\documentclass[10 pt, journal]{IEEEtran}

\makeatletter
\def\ps@headings{%
	\def\@oddhead{\mbox{}\scriptsize\rightmark \hfil \thepage}%
	
	\def\@evenhead{\scriptsize\thepage \hfil \leftmark\mbox{}}%
	
	\def\@oddfoot{}%
	
	\def\@evenfoot{}}
	
\makeatother

\usepackage[hidelinks]{hyperref}
\usepackage{cite}
\usepackage[table]{xcolor}
\usepackage{float}  
\usepackage{epsfig,bm,amsmath,amssymb,graphicx,setspace}
\usepackage[numbers,sort&compress]{natbib}
\usepackage{color,placeins}
\usepackage{multirow}
\usepackage{algorithm}
\usepackage{algpseudocode}
\usepackage{colortbl}
\usepackage{color}
\usepackage{dblfloatfix}
\usepackage{turnstile}
\usepackage{multicol}
\usepackage{array}

\usepackage{enumerate}
\usepackage{turnstile}
\usepackage{subcaption}
\usepackage{arydshln,paralist}
\usepackage{soul,tabularx,booktabs}
\usepackage{csquotes}
\usepackage{multirow}
\usepackage{adjustbox}
\usepackage{float}

\usepackage{xcolor}
\usepackage{verbatim}
\usepackage[colorinlistoftodos]{todonotes} 
\usepackage{rotating}
\definecolor{usethiscolorhere}{rgb}{0.86666,0.78431,0.78431}
\usepackage{tikz}
\usetikzlibrary{mindmap}
\usetikzlibrary{calc,positioning}
\usepackage{lipsum,adjustbox}
\usetikzlibrary{decorations.pathmorphing}

\usepackage{pifont}

\makeatother

\usepackage{amsmath}
\begin{document}

\title{FedSecureFormer: A Fast, Federated and Secure Transformer Framework for Lightweight Intrusion Detection in Connected and Autonomous Vehicles}

\author{Devika Sathyan, Vishnu Hari, Pratik Narang,~\IEEEmembership{Senior Member, ~IEEE}, and Tejasvi Alladi, ~\IEEEmembership{Senior Member, ~IEEE}, F. Richard Yu,~\IEEEmembership{Fellow,~IEEE}

\thanks{Devika Sathyan, Vishnu Hari, Pratik Narang and Tejasvi Alladi are with the Department of Computer Science and Information Systems, BITS Pilani, Pilani Campus, 333031, India.  (e-mail: p20210024@pilani.bits-pilani.ac.in; f20220094@pilani.bits-pilani.ac.in; pratik.narang@pilani.bits-pilani.ac.in; tejasvi.alladi@pilani.bits-pilani.ac.in).}

\thanks{F. Richard Yu is with the Department of Systems and Computer Engineering, Carleton University, Ottawa, ON K1S 5B6, Canada. (e-mail: richard.yu@carleton.ca).}}

\maketitle{}

\begin{abstract}
The era of Connected and Autonomous Vehicles (CAVs) reflects a significant milestone in transportation, improving safety, efficiency, and intelligent navigation. However, the growing reliance on real-time communication, constant connectivity, and autonomous decision-making raises serious cybersecurity concerns, particularly in environments with limited resources. This highlights the need for security solutions that are efficient, lightweight, and suitable for deployment in real-world vehicular environments. In this work, we introduce FedSecureFormer, a lightweight transformer-based model with 1.7 million parameters, significantly smaller than most encoder-only transformer architectures. The model is designed for efficient and accurate cyber attack detection, achieving a classification accuracy of 93.69\% across 19 attack types, and attaining improved performance on ten attack classes, outperforming several state-of-the-art (SOTA) methods.
To assess its practical viability, we implemented the model within a Federated Learning (FL) setup using the FedAvg aggregation strategy. We also incorporated differential privacy to enhance data protection. For testing its generalization to unseen attacks, we used a histogram-guided GAN with LSTM and attention modules to generate unseen data, achieving 88\% detection accuracy. Notably, the model reached an inference time of 3.7775 milliseconds per vehicle on Jetson Nano, making it roughly 100x faster than existing SOTA models. These results position FedSecureFormer as a fast, scalable, and privacy-preserving solution for securing future intelligent transport systems.

\end{abstract}


\section{Introduction}

The National Association of Counties (NACo) stated that Connected Vehicles (CV) are  \begin{em}``Vehicles that can communicate with other vehicles,  infrastructure, and devices through wireless network technology such as Wi-Fi and radio".\end{em} According to the US Department of Transportation, CVs utilize Vehicle-to-Everything (V2X) \cite{alladi2021artificial} communication to improve road safety, enable real-time data exchange, and ensure seamless interaction within highly dynamic vehicular networks by facilitating the exchange of critical information between vehicles, infrastructure, and networks \cite{devika2024vadgan}. Additionally, NACo defines Automated Vehicles (AV) (also known as driverless cars) as  \begin{em}``Vehicles equipped with technology that enables them to operate with little to no human assistance.\end{em}" AVs rely on integrated systems such as LiDAR, radar, cameras, and GPS \cite{multistage2024} for safe navigation. The integration of Connected and Autonomous Vehicles (CAVs) \cite{chougule2022multibranch} represents a significant advancement in transportation, with early milestones including Google's self-driving car project, now known as Waymo. Intelligent Transportation Systems (ITS) further enhance traffic safety and efficiency through technologies like Artificial Intelligence, sensor networks, and V2X, particularly using IEEE 802.11p \cite{chougule2022multibranch} for vehicular communication.

Despite their advantages, CAVs are vulnerable to a wide range of cybersecurity threats due to their reliance on interconnected systems, real-time data exchange, and wireless communication technologies \cite{alladi2021artificial}. These include Denial-of-Service (DoS) attacks, Sybil attacks, spoofing, and other similar attacks.  Vulnerabilities often stem from communication protocols, such as Vehicle-to-Everything (V2X) and Controller Area Network (CAN) \cite{CAVIDS}, as well as onboard components like LiDAR and GPS, due to the absence of encryption and authentication mechanisms.

Ensuring robust security measures has become essential with the increasing complexity of cyber threats in CAVs. This has led to the development of Intrusion Detection Systems (IDS) \cite{devika2024vadgan}. IDS solutions employ techniques such as machine learning \cite{slama2022comparative}, deep learning \cite{chougule2022multibranch}, and generative AI \cite{devika2024vadgan} to detect potential attacks in real-time. As cyberattacks become more sophisticated, the role of IDS in reinforcing vehicular cybersecurity continues to gain critical importance. However, these models face severe drawbacks due to data imbalance problems, long-range dependencies in network traffic, and data poisoning. Unlike traditional models, transformers utilize self-attention mechanisms to capture complex dependencies and patterns within sequential data from network traffic. 

Another key challenge is that existing IDS research for CAVs often prioritizes detection accuracy, with less emphasis on real-time response \cite{CAVIDS}. However, in dynamic environments, the ability to detect and respond to threats promptly is just as critical as ensuring vehicle safety. Centralized models often struggle with non-IID (non-Independent and Identically Distributed) data and pose a single point of failure, making them vulnerable to attacks. This limitation has led to the adoption of Federated Learning (FL) in the CAV domain \cite{flid}, which enhances robustness by enabling distributed learning across multiple clients while improving adaptability to non-IID data through a client-server architecture.

To address the limitations identified in the existing literature, we developed a fast and lightweight transformer-based model designed explicitly for intrusion detection in CAV environments. This direction has received limited attention so far. Although integrating FL with privacy-preserving mechanisms such as Differential Privacy (DP) or Homomorphic Encryption offers enhanced security, this approach remains underexplored in CAV-focused IDS research. In addition, most existing studies have overlooked the inference time in resource-constrained environments, even though a fast response is crucial for timely threat detection and decision-making.

The main contributions of this work are as follows:

\begin{enumerate}[i] \item \textbf{Lightweight Transformer Architecture}: We propose FedSecureFormer, a compact transformer model with only 1.7 million parameters, tailored for intrusion detection in Connected and Autonomous Vehicles (CAVs), achieving improved performance across ten out of 19 cyberattack classes.

\item \textbf{Detection of Unseen Attacks}: To evaluate generalization, we tested the model on unseen adversarial sequences generated by a histogram-based GAN with LSTM and attention layers. FedSecureFormer achieved an 88\% detection rate, demonstrating robust adaptability.

\item \textbf{Federated and Privacy Preserving Deployment}: The model was deployed in a Federated Learning setup with FedAvg, showing less than 1.03\% performance drop. With differential privacy, it maintained a 4.04\% accuracy reduction across 20 clients.

\item \textbf{Real Time Inference Capability}: FedSecureFormer achieved an inference time of 3.7775 milliseconds per vehicle on Jetson Nano, making it nearly 100 times faster than existing models and well-suited for deployment in resource-constrained Intelligent Transport Systems. \end{enumerate}

The remainder of the paper is organized as follows. Section II reviews related work, while Section III outlines the background, including the CAV framework and a taxonomy of vehicular misbehaviour. Section IV details the proposed FedSecureFormer architecture in both centralized and federated settings. Section V describes the experimental setup, followed by results and analysis in Section VI. Finally, Section VII concludes the study.

\section{Literature Survey}
\label{sec:literature}
Our literature review is structured as follows: We begin by delving into the IDS framework within the context of CAVs and then exploring transformer-based approaches for intrusion detection. Next, we examine real-time response capabilities in resource-constrained environments, followed by techniques for detecting previously unseen attacks, and conclude with a review of FL frameworks applied to IDS in CAV scenarios.

Intrusion detection within CAV environments involves identifying and categorizing misbehaviours exhibited by participant vehicles. Sharma et al. \cite{hongliu20} calculated plausibility scores related to the location and movement of vehicles, employing a confidence interval-based technique to identify physically improbable behaviours within the VeReMi Dataset ~\cite{veremi-og}. Whereas Sedar et al. \cite{veremi-rl} considered the intrusion detection task a reinforcement learning problem, implementing an LSTM-based Q-network for classification purposes. Moreover, Alladi et al. \cite{alladi2021artificial} deployed hybrid CNN-LSTM models on edge devices to facilitate rapid and effective detection of intrusions.

Recent advancements have highlighted the effectiveness of transformer-based architectures in IDS for CAVs. Wang et al. \cite{hierarchichal-vector-transformer} demonstrated strong detection performance using hierarchical vector transformers, which employed distinct self-attention mechanisms to capture inter-vehicular interactions and temporal dependencies separately. Similarly, Guan et al. \cite{transformers-classification} improved detection accuracy by integrating Principal Component Analysis (PCA) with multi-head self-attention, illustrating the value of hybrid transformer-based approaches. Additionally, Mundra et al. \cite{mundra2025decentralized} proposed a lightweight transformer architecture in a decentralized framework by distributing the processing tasks and feature selection across multiple vehicles.

Computational latency and response time are critical considerations in CAV scenarios, where real-time feedback is essential to ensure safety and operational efficiency. Kumar et. al \cite{CAVIDS} reported promising results on the CarChallenge dataset \cite{kang2021car} using a Raspberry Pi, achieving inference times below 50 microseconds. In contrast, Alladi et al. \cite{alladi2021artificial} documented a significantly higher average inference time across multiple models, at 234.505 milliseconds per vehicle, on a similar hardware setup, highlighting the wide variation in performance across different IDS implementations. Meanwhile, Mundra et al. \cite{mundra2025decentralized} reported an average detection time of 22 to 25 milliseconds, further emphasizing the need for optimized, low-latency solutions in resource-constrained environments. 

The effectiveness of an IDS is significantly strengthened by its ability to detect unseen attacks. Althunayyan et al. \cite{multistage2024} proposed a two-stage architecture comprising an Artificial Neural Network (ANN) for detecting and classifying known attacks, and an LSTM autoencoder for identifying previously unseen attacks. Lin et al. \cite{phade} addressed visual spoofing attacks using a multimodal sensor fusion approach, introducing a multi-head self-attention-based fusion module to enhance robustness. Similarly, Fan et al. \cite{fan2024auto} developed a two-layer filtering mechanism that employed a One-Class SVM to detect unknown attacks and dynamically update the attack recognizer, which consists of a 1-D CNN augmented with a self-attention mechanism.

Liu et al. \cite{liu2021blockchain} proposed a decentralized FL framework that integrates a trust-based consensus mechanism using blockchain, where model updates are aggregated across multiple Roadside Units (RSUs), and evaluated using the KDD99 dataset. Bhavsar et al. \cite{bhavsar2024fl} demonstrated a practical FL setup using Raspberry Pi clients and a Jetson Xavier server, evaluating NSL-KDD and Car-Hacking datasets, with FedAvg and FedYogi aggregation strategies. Additionally, Al-Hawawreh et al. \cite{al2023federated} introduced a distributed FL approach based on Augmented Lagrangian optimization, enabling each client to aggregate updates from its neighbouring clients, thereby enhancing collaboration in decentralized settings.

While prior studies such as \cite{alladi2021artificial}, \cite{hongliu20}, \cite{phade}, \cite{fan2024auto}, \cite{liu2021blockchain}, \cite{bhavsar2024fl}, and \cite{al2023federated} have demonstrated strong detection performance, they fall short in addressing several essential aspects for a practical inter-vehicular IDS. Specifically, \cite{alladi2021artificial} has limited its evaluation to a subset of attack classes, leaving a significant portion unexamined, whereas \cite{hongliu20} has utilized an older version of the VeReMi Extension dataset \cite{veremi-og}, which comparatively has fewer attack types. Although \cite{fan2024auto} addresses the detection of previously unseen attacks, it is based on an update mechanism that is prone to misclassifying known attack types, as acknowledged by the authors. Similarly, \cite{phade} introduced artificial modalities, but the work focused only on phantom spoofing attacks. In contrast, \cite{liu2021blockchain}, \cite{bhavsar2024fl}, and \cite{al2023federated} utilized FL; however, \cite{liu2021blockchain} used the obsolete KDD99 dataset, \cite{bhavsar2024fl} restricted scalability to only four clients, and \cite{al2023federated} relied on a centralized model with no consideration for deployment in resource-constrained environments.

Addressing these limitations collectively, we propose a lightweight IDS architecture optimized for deployment in resource-constrained environments. Despite the proven effectiveness of transformers in various security applications, their application within the vehicular domain, particularly in a protected FL setup, remains limited. This gap in current research motivates the development of a robust, scalable, and privacy-aware IDS tailored to the unique challenges of the CAV IDS.

\section{Preliminary Background}
\label{sec:background}
\subsubsection{CAV Scenario}
\captionsetup[figure]{skip=0.5pt}
\begin{figure}
    \centering
    \includegraphics[width=0.3\textwidth]{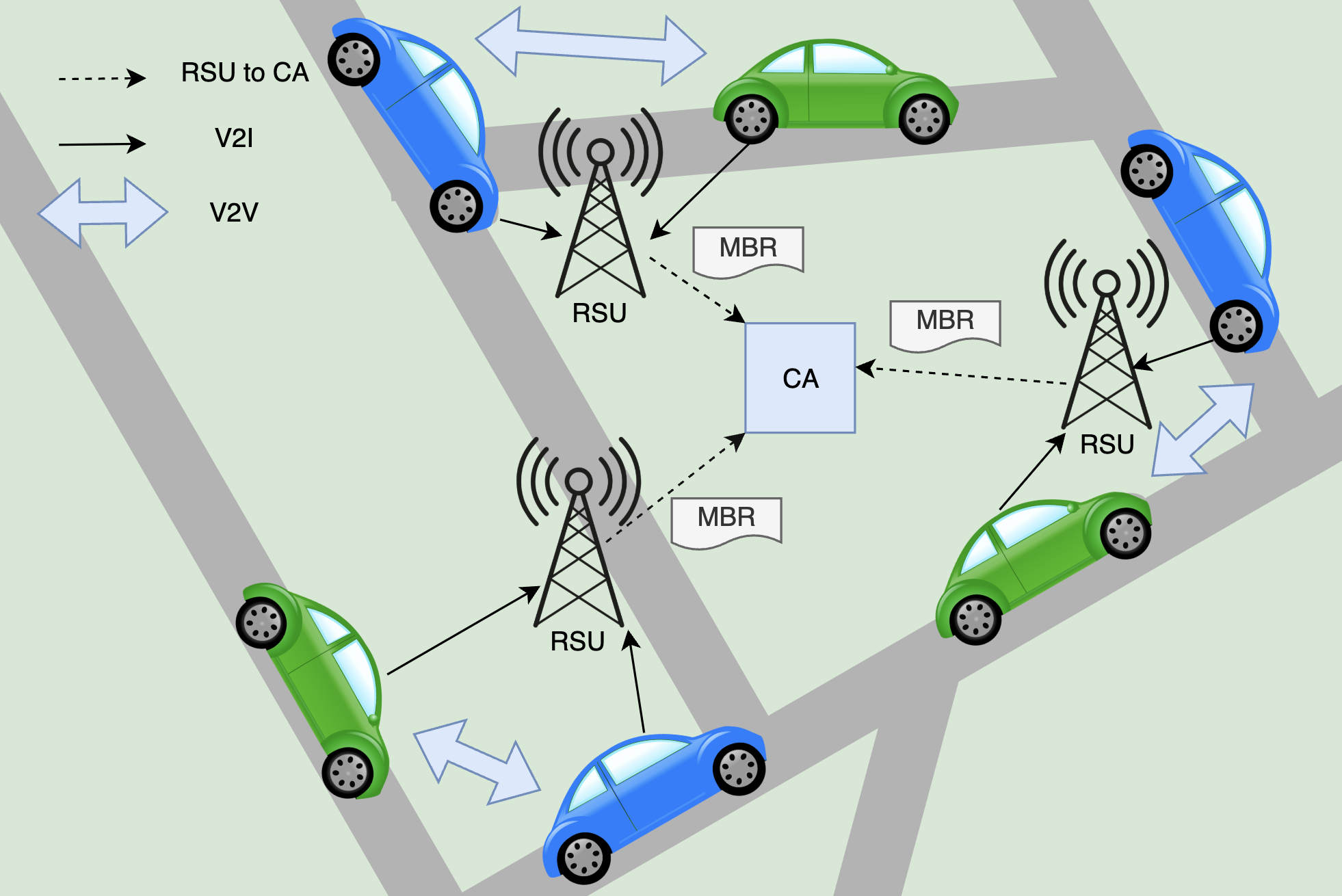}
    \caption{CAV Scenario}
    \label{fig:cav-scenario}
\end{figure}
A typical Connected and Autonomous Vehicle (CAV) scenario, as illustrated in Fig. \ref{fig:cav-scenario}, features Vehicle-to-Everything (V2X) communication, where a vehicle interacts with other vehicles and fixed infrastructure, such as traffic signals and road signage. Each vehicle has an Onboard Unit (OBU) that processes and fuses data from GPS and other onboard sensors. This information is transmitted to nearby RSUs, which are managed by the region's Control Authority (CA). RSUs are equipped with built-in mechanisms for detecting and classifying misbehaviour with the installed IDS systems. When anomalous activity is identified, the RSU generates a MisBehavior Report (MBR) and forwards it to the CA. The CA then evaluates the report and decides on appropriate actions, such as issuing alerts to other vehicles or revoking the certificate of the offending vehicle.

\subsubsection{Dataset}In this paper, we use the VeReMi Extension Dataset~\cite{veremi-extension}, an enhancement of the original VeReMi dataset~\cite{veremi-og}, which is a synthetically generated dataset incorporating a realistic sensor error model to reflect real-world scenarios better. The attack taxonomy is detailed in~\cite{devika2024vadgan}. We have used nine features from the dataset, namely, timestamp (the time at which the message was sent) and the motion coordinates of the vehicle in x and y-coordinates, such as position, speed, acceleration\textcolor{blue}{,} and heading. 
The attack classes are as follows:
A(0) - Normal, A(1) Constant position, A(2) - Constant position offset, A(3) - Random position, A(4) - Random position offset, A(5) - Constant speed, A(6) – Constant speed offset, A(7) – Random speed, A(8) – Random speed offset, A(9) – Eventual Stop, A(10) - Disruptive, A(11) - Data replay, A(12) - Delayed messages, A(13) - DoS, A(14) - DoS random, A(15) - DoS Disruptive, A(16) - Data replay sybil, A(17) - Traffic congestion sybil, A(18) - DoS random sybil and A(19) - DoS disruptive sybil.

\subsubsection{Transformers}
The Transformers, first introduced in \cite{attention-is-all-you-need}, are deep neural networks that process sequential data effectively. Transformers replace conventional sequential models, such as RNNs and LSTMs, with parallel processing of data sequences. Parallel processing in transformers is facilitated by a self-attention mechanism, which enables the capture of long-range dependencies in data sequences by learning the relationships between different data elements within the sequence. While initially introduced for Natural Language Processing (NLP) tasks, transformers have shown promising results in a wide range of domains, including temporal sequence modelling tasks \cite{transformers-classification,hierarchichal-vector-transformer}.

\subsubsection{Federated Learning}
Federated Learning (FL)~\cite{fedavg} is a machine learning framework that can collaboratively train a shared model without exchanging data across different devices. Each round of FL involves training the model locally on clients (can be edge devices) and then sharing the model updates with the server. The final recombination of model updates is carried out using strategies such as the Federated Average Strategy (FedAvg) \cite{fedavg}, a weight averaging technique, or the Federated Proximal Strategy (FedProx) \cite{fedprox}, an iteration on FedAvg designed to improve its performance in heterogeneous environments. In this work, we have utilized FedAvg and FedProx strategies because, as stated in the existing literature \cite{herlambang2025federated,karimireddy2020scaffold}, both strategies exhibit similar complexities and convergence behaviour in homogeneous environments. 
\captionsetup[figure]{skip=0.5pt}
\begin{figure*}
    \centering
    \includegraphics[width=0.8\textwidth]{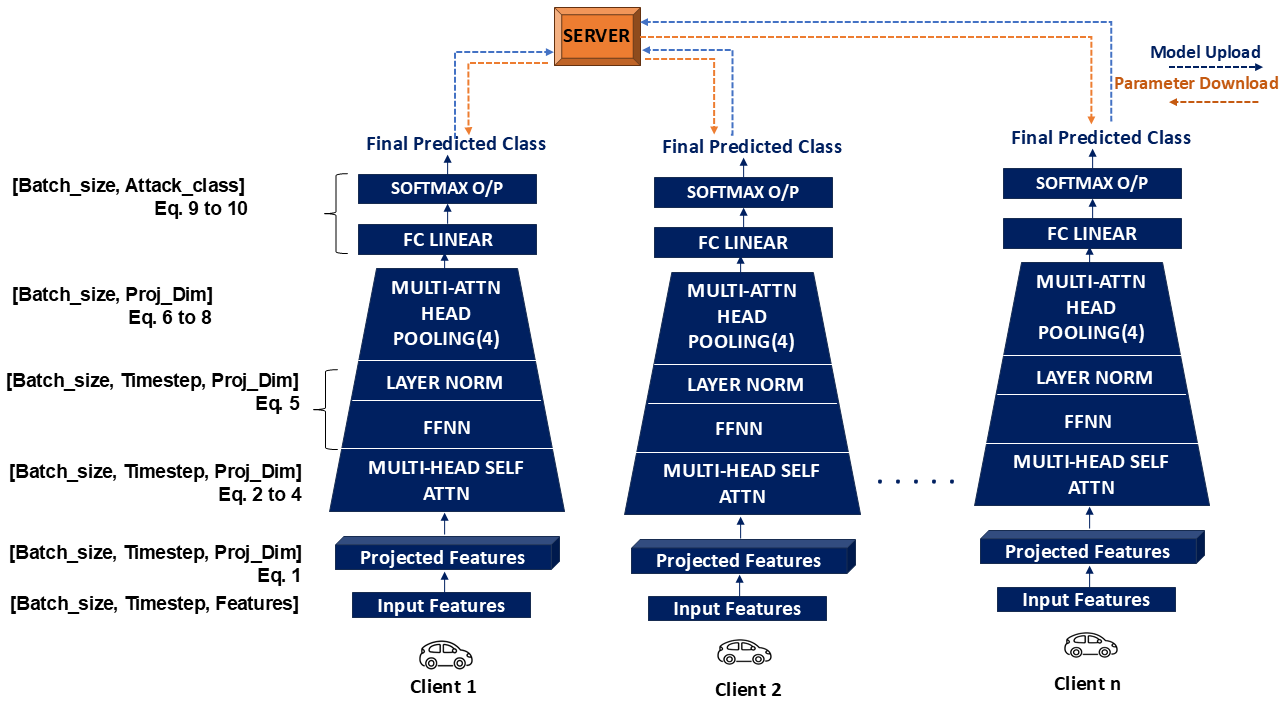}
    \caption{FedSecureFormer architecture based on a 6-layer encoder-only transformer model, designed for efficient intrusion detection in CAVs, illustrated within a Federated Learning setup involving $n$ clients.}
    \label{fig:FL}
\end{figure*}
More recent studies \cite{flid, multistage2024} demonstrate that FL can effectively enable a decentralized approach while also addressing privacy preservation.

\section{Proposed Intrusion Detection Framework}
\label{sec:proposed}
The proposed FedSecureFormer architecture is illustrated in Fig. \ref{fig:FL}. FedSecureFormer is a transformer architecture specifically designed for IDS in the CAV scenario. The model was implemented in both centralized and distributed settings.

\subsection{Proposed FedSecureFormer Transformer}
 
As shown in Fig. \ref{fig:FL}, the transformer architecture employed in our work follows an encoder-only design consisting of six stacked encoder layers. The input is first projected into an embedding space using a linear layer, followed by the addition of learnable absolute positional encodings to incorporate temporal context. Each encoder layer contains multi-head self-attention with two attention heads, each having its own Query, Key, and Value projections. This is followed by layer normalization and a two-layer Feed Forward Network (FFNN) with ReLU activation. Residual connections are applied after the attention and FFNN blocks to improve gradient flow and model stability. Post encoding, the output is passed through a multi-head attention pooling layer with four heads. This layer uses a multi-query strategy to share Keys and Values across heads. In contrast, each head maintains a unique learned Query vector, resulting in concatenated context vectors to form a pooled representation. Finally, this is passed through a fully connected linear layer and softmax to generate the final probability distribution over the target classes. The mathematical formulas used are as follows:

\begin{equation}
Z = \text{Stack}(\mathbf{z}_1, \dots, \mathbf{z}_t), \quad
\text{where } \mathbf{z}_t = W_{\text{proj}} \cdot \mathbf{x}_t + \mathbf{b}_{\text{proj}} + \mathbf{p}_t
\end{equation}

\begin{equation}
Q_i = Z W^Q_i, \quad K_i = Z W^K_i, \quad V_i = Z W^V_i
\end{equation}
\begin{equation}
\text{head}_i = \text{softmax} \left( \frac{Q_i K_i^\top}{\sqrt{d_k}} \right) V_i
\end{equation}
\begin{equation}
\text{MHSA}(Z) = \text{Concat}(\text{head}_1, \dots, \text{head}_H) \cdot W_O
\end{equation}

\begin{equation}
\begin{aligned}
Z' &= \text{LayerNorm}\left(Z + \text{FFN}(Z)\right) \\
\text{where} \quad \text{FFN}(Z) &= \text{ReLU}(Z W_1 + b_1) W_2 + b_2
\end{aligned}
\end{equation}

\begin{equation}
q_j = u_j w^q_j, \quad k = Z' w^K, \quad v = Z' w^V
\end{equation}
\begin{equation}
\text{context}_j = \text{softmax} \left( \frac{q_j k^\top}{\sqrt{d_k}} \right) v
\end{equation}
\begin{equation}
\text{pooled} = W_{\text{pool}} \cdot \text{Concat}(\text{context}_1, \dots, \text{context}_{H_{\text{pool}}})
\end{equation}
\begin{equation}
\mathbf{y} = W_{\text{cls}} \cdot \text{pooled} + \mathbf{b}_{\text{cls}}
\end{equation}
\begin{equation}
\hat{\mathbf{p}} = \text{softmax}(\mathbf{y}), \quad
\hat{c} = \arg\max(\hat{\mathbf{p}})
\end{equation}

\begin{equation}
\mathcal{L}_{\text{SmoothL1}}(\hat{p}, y) = 
\begin{cases}
\frac{1}{2} (\hat{p}_k - y_k)^2, & \text{if } |\hat{p}_k - y_k| < 1 \\
|\hat{p}_k - y_k| - \frac{1}{2}, & \text{otherwise}
\end{cases}
\end{equation}

Here, the input feature vector $\mathbf{x}_t$ at timestep $t$ is projected to $d_\text{dim}$ using the weight matrix $W_{\text{proj}}$, bias $\mathbf{b}_{\text{proj}}$, and added to a learnable positional encoding $\mathbf{p}_t$ to produce the embedded vector $\mathbf{z}_t$. The complete input sequence is represented as $Z$. In the self-attention mechanism, $Q_i$, $K_i$, and $V_i$ are the Query, Key, and Value matrices for attention head $i$, each of dimension $d_k$. The $d_k$ is calculated by dividing $d_\text{dim}$
by the number of attention heads $H$. The outputs from all heads are combined via an output projection $W_O$. Next, the output is fed to an FFN, followed by a layer normalization layer, resulting in Z'. Each head $j$ uses a learned query vector $q_j$ (utilizing a learnable vector $u_j$) in the multi-head pooling stage, while key $k$ and value $v$ from encoder outputs are shared across heads. The  $\text{context}_j$ learnt $q_j$, $k$, and $v$ are concatenated and projected using $W_{\text{pool}}$ to yield the final pooled vector, utilizing the pooling heads $H_\text{pool}$. For classification, the pooled vector is passed through a fully connected layer with weights $W_{\text{cls}}$ and a bias vector $\mathbf{b}_{\text{cls}}$ to produce logits $\mathbf{y}$. The predicted probabilities $\hat{\mathbf{p}}$ are obtained via softmax, and the final predicted class $\hat{c}$ corresponds to the class with the highest probability. Finally, the loss is computed via smooth L1 loss using the predicted $\hat{\mathbf{p}}_k$ and actual probabilities $y_k$.

\subsection{Histogram-aware Attention GAN (Hist-AttnGAN)}
\begin{figure}

    \centering
    \includegraphics[width=0.4\textwidth]{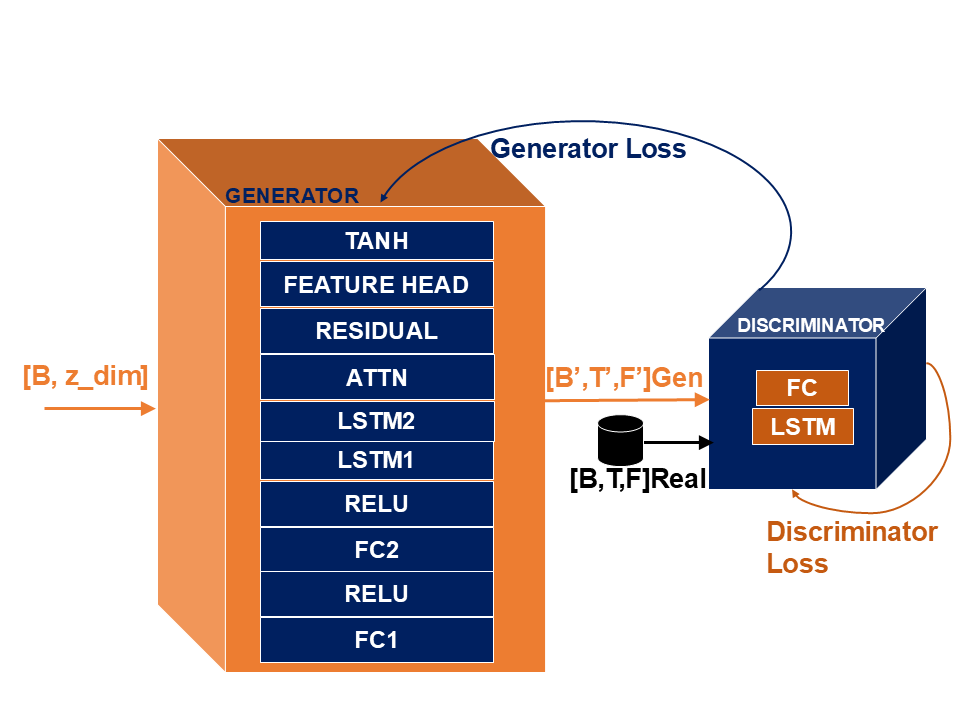}
    \caption{His-AttnGAN architecture utilized for generating unseen data.}
    \label{fig:hisgan}
\end{figure}

To detect unseen attacks, we generated realistic samples using a histogram-guided GAN as depicted in Fig. \ref{fig:hisgan}. The generator component of this GAN comprises a two-layer LSTM architecture enhanced with multi-head self-attention utilizing four attention heads. Each timestep's output is subsequently passed through nine parallel feature projection heads—one for each feature—each composed of a small Multi-Layer Perceptron (MLP) consisting of combinations of Dense and LeakyReLU layers to independently generate feature values. The discriminator is constructed using a single-layer LSTM. During training, the generator receives input noise shaped \([Batch\_size, Noise\_Dimension ]\) and produces outputs of shape \([Batch\_size', Timesteps', Features']\). Both the generated data and the original training data, formatted identically as \([Batch\_size, Timesteps, Features]\), are provided as inputs to the discriminator. The model training is guided by two primary objective functions: a discriminator loss, calculated using the Wasserstein distance coupled with a gradient penalty, and a generator loss that incorporates both adversarial loss and a histogram-based distribution loss. 

\captionsetup[figure]{skip=0.8pt}

\begin{equation}
\mathcal{L}_G = -\mathbb{E}_{\tilde{x} \sim \mathbb{P}_{\text{fake}}} \left[ D(\tilde{x}) \right] + \lambda_{\text{hist}} \cdot \sum_{i=1}^{F} \left\| \text{CDF}^{(i)}_{\text{real}} - \text{CDF}^{(i)}_{\text{fake}} \right\|_1
\end{equation}

\begin{equation}
\mathcal{L}_D = -\mathbb{E}_{x \sim \mathbb{P}_{\text{real}}} \left[ D(x) \right] + \mathbb{E}_{\tilde{x} \sim \mathbb{P}_{\text{fake}}} \left[ D(\tilde{x}) \right] + \text{GP}
\end{equation}

\begin{equation}
\text{GP} = \lambda_{\text{gp}} \cdot \mathbb{E}_{\hat{x} \sim \mathbb{P}_{\hat{x}}} \left[ \left( \left\| \nabla_{\hat{x}} D(\hat{x}) \right\|_2 - 1 \right)^2 \right]
\end{equation}

The generator loss $\mathcal{L}_G$ is calculated as the adversarial loss added with histogram loss, which is the Cumulative Distribution Functions' (CDF) L1 loss of the $i^{th}$ real and fake features, respectively. The histograms are first constructed for each $i^{th}$ real and fake feature and then normalized into probability distributions. These distributions are then converted into CDFs, which are compared to measure the difference between the real and generated sequences. The $\lambda_{\text{hist}}$ is a weighting factor for histogram loss. To maintain equal weightage for both adversarial loss and histogram loss, $\lambda_{\text{hist}}$ was assigned a value of 1. $\mathcal{L}_D$ (discriminator loss) is the sum of Wasserstein's loss and gradient penalty ($GP$). $GP$ is calculated using the L2 norm of the discriminator gradient, along with a penalty for deviation from the unit norm case. $\lambda_{\text{gp}}$ is the Gradient penalty coefficient set to 10. 

\subsection{Federated Learning Framework with Differential Privacy}
Our FL implementation aligns closely with the vanilla FL algorithm \cite{flid}. To deal with the class imbalance issue while splitting the data between multiple clients, we have computed the class weights ($\mathcal{W}_c$) and used them as a parameter for the weighted cross-entropy loss $(\mathcal{L}_{\mathrm{WCE}}
)$. The aggregation of model parameters from different clients is carried out using the FedAvg and FedProx strategies, with FedAvg yielding the best results. 

\begin{equation}
\mathcal{L}_{\mathrm{WCE}} = -\sum_{i=1}^{C} w_i \cdot y_i \cdot \log(\hat{y}_i)
\end{equation}

\begin{equation}
w_{\text{global}} = \frac{1}{N} \sum_{i=1}^{K} n_i \cdot w_i
\label{eq:2}
\end{equation}

\begin{equation}
\text{N} = \sum_{i=1}^{K} n_i
\end{equation}

\begin{equation}
\text{M}_{\text{global}} = \frac{1}{N} \sum_{i=1}^{K} n_i \cdot \text{metric}_i
\label{eq:1}
\end{equation}

\begin{equation}
\mathcal{L}_{\mathrm{WCE}}(\theta)= \mathcal{L}(\theta) + \frac{\mu}{2} \| \theta - \theta^{(t)} \|^2
\label{eq:3}
\end{equation}

Here $C$, $y_i$, $\hat{y}_i$\textcolor{blue}{,} and $w_i$ represent the number of attack classes, true class label, predicted class probability, and class weight for each class $i$, respectively. Whereas in the equation used to compute the aggregation strategy, we have the following terms $K$, $n_i$, $w_i$, $N$, which represent the total number of clients, the number of training samples for each client, the local model parameter for each client $c$, and the total number of training samples across all clients. As shown in eq. \ref{eq:1}, we calculated the average of all performance measures across all clients. In eq. \ref{eq:3}, we calculated the aggregation using FedProx strategy, utilizing the weighted cross-entropy loss function ($\mathcal{L}_{\mathrm{WCE}}(\theta)$), current local model parameters ( $\theta$), global model parameters received from the server before local training ($\theta^{(t)}$), and the FedProx constant (proximal term weight, $\mu$). 

To ensure privacy-preserving training in the federated environment, gradient updates are adjusted by applying gradient clipping and adding noise, as described below:

\begin{equation}
    \tilde{g}_i = \frac{g_i}{\max\left(1, \frac{\|g_i\|_2}{C_{\text{clip}}} \right)}
\end{equation}


\begin{equation}
\hat{g} = \frac{1}{B} \left( \sum_{i=1}^{B} \tilde{g}_i + \mathcal{N}(0,\, \sigma^{2} C_{\text{clip}}^{2} I) \right)
\end{equation}

\noindent
In the above equations, \(g_i\) denotes the original gradient of the \(i\)-th model parameter, while \(\tilde{g}_i\) represents the clipped gradient. The clipping threshold $C_{clip}$ restricts individual gradients to a specified maximum \(\ell_2\) norm. After gradient clipping, Gaussian noise \(\mathcal{N}(0, \sigma^2 C_{\text{clip}}^2I)\) is introduced to the averaged gradient (per batch \(B\)), with \(\sigma\) acting as the noise multiplier that controls the magnitude of the privacy-preserving noise. The resultant noisy gradient \(\hat{g}_i\) guarantees DP by ensuring that gradient updates remain statistically indistinguishable, even when a single training sample is altered or removed. The privacy guarantee is formally denoted by two privacy budget parameters $(\epsilon,\delta)$, where $\epsilon$ refers to the maximum allowable privacy loss and $\delta$ refers to the probability of privacy failure with respect to $\epsilon$, without compromising the model's utility (performance) \cite{de2022unlocking}.

 The DP accountant utilized in our work is Rényi DP (RDP) \cite{mironov2017renyi, wang2019subsampled}, and the privacy budget parameters $(\epsilon,\delta)$ are calculated using the mechanism as described in \cite{wang2019subsampled,mironov2019r}. 

\begin{equation}
\varepsilon(\alpha) = \frac{\alpha C_{\text{clip}}^{2}}{2\sigma^{2}}
\end{equation}

\begin{equation}
\varepsilon'(\alpha) = \frac{1}{\alpha - 1} 
\log \mathbb{E}_{k \sim \mathcal{B}(\alpha,\gamma)}
\left[ \exp\left( (\alpha - 1)\cdot \varepsilon_k \right) \right]
\end{equation}

\begin{equation}
\varepsilon_T(\alpha) = T \cdot \varepsilon'(\alpha)
\end{equation}

\begin{equation}
\label{eq.final}
\varepsilon(\delta) = 
\min_{\alpha > 1} \left( 
\varepsilon_T(\alpha) + 
\frac{\log(1/\delta)}{\alpha - 1} 
\right)
\end{equation}

Here, $\varepsilon(\alpha)$ is the RDP parameter with sensitivity $C_\text{clip}$. $\alpha$ corresponds to the order of Rényi divergence, a tunable parameter with value greater than 1, and we have used values typically ranging from 1.01 to 64. $\varepsilon'(\alpha)$ provides the RDP after subsampling, where $\gamma$ corresponds to the sampling probability, $ \mathcal{B}(\alpha,\gamma)$ is the binomial distribution used in subsampling RDP analysis, and $\varepsilon_k$ corresponds to RDP parameter for k-fold group. $\varepsilon_T(\alpha)$ corresponds to the composition rule for $T$ local rounds. The final eq. \ref{eq.final} shows the conversion from RDP to $(\epsilon,\delta)$-DP guarantee. The algorithm for the proposed FedSecureFormer is outlined as
follows in Algorithms \ref{alg:fedsecureformer-inference} and \ref{alg:fl-dp-secureformer}.
\begin{algorithm}[t]
\caption{FedSecureFormer Algorithm}
\label{alg:fedsecureformer-inference}
\begin{algorithmic}[1]
\State \textbf{Input:} $X \in \mathbb{R}^{T \times F}$ (input sequence), model weights $\Theta = \{W_{\text{proj}}, \mathbf{b}_{\text{proj}}, \mathbf{p}_t, W_i^Q, W_i^K, W_i^V, W_O, W_{\text{pool}}, W_{\text{cls}}, \mathbf{b}_{\text{cls}}, q_i\}$, shared keys $K$, values $V$
\State \textbf{Output:} Predicted class $\hat{c}$, probability vector $\hat{\mathbf{p}}$
\State \textbf{Hyperparameters:} $T=20$ (time steps), $F=9$ (features), $H=2$ (attention heads), $H_{\text{pool}}=4$ (pooling heads), $d_k=32$ (attention head dimension), $d_{\text{dim}} = 64$ (projection dimension), $C=20$ (number of classes), 
\State $d_k = \frac{d_{\text{dim}}}{\text{H}}$

\For{$t = 1$ to $T$}
    \State $\mathbf{z}_t = W_{\text{proj}} \cdot \mathbf{x}_t + \mathbf{b}_{\text{proj}} + \mathbf{p}_t$
\EndFor
\State $Z = \text{Stack}(\mathbf{z}_1, \dots, \mathbf{z}_T)$

\For{$i = 1$ to $H$}
    \State $Q_i = Z W_i^Q$
    \State $K_i = Z W_i^K$
    \State $V_i = Z W_i^V$
    \State $\text{head}_i = \text{softmax} \left( \frac{Q_i K_i^\top}{\sqrt{d_k}} \right) V_i$
\EndFor
\State $\text{MHSA}(Z) = \text{Concat}(\text{head}_1, \dots, \text{head}_H) \cdot W_O$
\State  $\text{FFN}(Z) = \text{ReLU}(Z W_1 + b_1) W_2 + b_2$
\State  $Z' = \text{LayerNorm}\left(Z + \text{FFN}(Z)\right)$

\For{$j = 1$ to $H_{\text{pool}}$}
    \State $q_j = u_j w_j^Q$
    \State $k = Z' w^K$
    \State $v = Z' w^V$
    \State $\text{context}_j = \text{softmax} \left( \frac{q_j k^\top}{\sqrt{d_k}} \right) \cdot v$
\EndFor
\State $\text{pooled} = W_{\text{pool}} \cdot \text{Concat}(\text{context}_1, \dots, \text{context}_{H_{\text{pool}}})$

\State $\mathbf{y} = W_{\text{cls}} \cdot \text{pooled} + \mathbf{b}_{\text{cls}}$
\State $\hat{\mathbf{p}} = \text{softmax}(\mathbf{y})$
\State $\hat{c} = \arg\max(\hat{\mathbf{p}})$

\State $\mathcal{L}_{\text{SmoothL1}}(\hat{p}_k, y_k) = 
\begin{cases}
\frac{1}{2} (\hat{p}_k - y_k)^2, & \text{if } |\hat{p}_k - y_k| < 1 \\
|\hat{p}_k - y_k| - \frac{1}{2}, & \text{otherwise}
\end{cases}$
\end{algorithmic}
\end{algorithm}

 \begin{algorithm}[t]
\caption{FedSecureFormer in FL-DP Algorithm}
\label{alg:fl-dp-secureformer}
\begin{algorithmic}[1]
\State \textbf{Input:} $D = \bigcup_{i=1}^{N} D_i$ (federated dataset including X and y features across $N$ clients), $R = 100$ (total rounds), $E = 1$ (epoch per round), $B$ (batch size), $\eta$ (learning rate), $\sigma$ (noise multiplier), $C_{clip}$
 (clipping norm), $\Theta_0$ (initial global weights), $T = 20$ (time steps), $F = 9$ (features)
\State \textbf{Output:} Final global model $\Theta_R$
\State Initialize FedSecureFormer weights $\Theta_0$ on the server
\For{$r = 1$ to $R$}
    \State Broadcast global weights $\Theta_{r-1}$ to all clients $i = 1$ to \hspace*{0.5cm} $N$
    \ForAll{client $i = 1$ to $N$ \textbf{in parallel}}
        \State Receive $\Theta_{r-1}$, set $\Theta_i \gets \Theta_{r-1}$
        \ForAll{batch $(X, y) \in D_i$}
            \State $\hat{y} \gets \text{FedSecureFormer}(X; \Theta_i)$
            \State $\mathcal{L}_{\text{WCE}} = - \sum_{c=1}^{C} w_c \cdot y_c \cdot \log(\hat{y}_c)$
            \State Compute gradients: $\nabla_{\theta_i} \gets \nabla \mathcal{L}_{\text{WCE}}$
            \State Clip gradients: $\nabla \theta_j \gets \nabla_{\theta_i} / \max(1, \frac{\|\nabla_{\theta_i}|_2}{C_{\text{clip}}})$
        { \State Add noise and average:  
\Statex \quad \hspace{3.0em} $\nabla \theta_j \gets 
\frac{1}{B} \left( \sum_{i=1}^B \theta_j + 
\mathcal{N}(0, \sigma^2 C_{\text{clip}}^2 I) \right)$}

            \State Update model: $\Theta_i \gets \Theta_i - \eta \cdot \nabla \theta_j$
        \EndFor
        \State Return $\Theta_i$ and metrics (Accuracy, F1, etc.)
    \EndFor
    \State Aggregate: $\Theta_r \gets \frac{1}{\sum n_i} \sum n_i \cdot \Theta_i$
\EndFor
\State \Return final model $\Theta_R$
\end{algorithmic}
\end{algorithm}
 
\section{Experimental Setup}
\label{sec:simulation}
This section provides details of the system specification, dataset specification\textcolor{blue}{,} and model hyperparameters used in this study. 

\subsection{System Specification}

All experiments were conducted on an NVIDIA RTX A6000 GPU within a software environment consisting of Python 3.10.12, PyTorch 2.6.0, CUDA 11.8\textcolor{blue}{,} and the Visual Studio Code (VSCode) IDE. FL experiments were implemented using the Flower framework (version 1.14.0). The proposed FedSecureFormer model was also deployed on a Jetson Nano running JetPack 4.6.6 to demonstrate real-time inference and evaluate performance in resource-constrained environments.

\subsection{Dataset Specification}

The dataset used in this study is the VeReMi Extension dataset \cite{veremi-extension}. Prior studies \cite{chougule2022multibranch,alladi2023deep} have inspired us to use the sliding window mechanism with a window size of 20, where each sample captures 20 consecutive timesteps with nine features per timestep, resulting in an input shape of [20,9] for each sample. A slide length of 10 was used, resulting in lower repetition and a higher number of sequences. Any sequence shorter than 20 was discarded. Each sequence is associated with one of 20 distinct classes (A(0) to A(19)), representing various types of attack categories. 
The dataset was labeled and sequentially generated, producing the following sequence distribution per class: A(0) - 165373, A(1) - 3804, A(2) - 3793, A(3) - 3821, A(4) - 3701, A(5) - 3623, A(6) - 3886, A(7) - 3662, A(8) - 3705, A(9) - 3727, A(10) - 3776, A(11) - 3870, A(12) - 3745, A(13) - 12564, A(14) - 12103, A(15) - 12370, A(16) - 16981, A(17) - 3876, A(18) - 8122, and A(19) - 7654. The dataset was split into a 70:15:15 ratio for training, validation, and testing, respectively, ensuring that the original class imbalance was preserved.

\subsection{Model Hyperparameters}

The model took approximately 6937.16 seconds to train for 100 epochs using the Adam optimizer (learning rate = 0.0003). A gradient penalty of 10 and a latent space dimension of 100 were used. The hyperparameter choices for our transformer architecture were selected through an extensive ablation study (Section \ref{sec:results}), resulting in six encoder layers, each equipped with two attention heads, and a multi-head attention pooling layer with four heads. In a centralized training, we set the projection dimension to 64, the batch size to 128, and utilized smooth L1 loss. For the FL setup, we have used a batch size of 64 and weighted cross-entropy loss to address the class imbalance introduced when splitting the data across multiple clients. The FL experiments utilized 20 clients for 100 rounds and one local epoch, with a clip norm of 5 and a noise multiplier of 0.001. For the His-AttnGAN, we considered using a batch size of 64 and a noise dimension of 128.

\begin{table}
\centering
\caption{Comparison of Model Architectures – Hybrid CNN-LSTM and Temporal Convolutional Network against various Performance Measures}
\footnotesize
\resizebox{0.8\columnwidth}{!}{%
\begin{tabular}{lcccc}
\toprule
\textbf{Model} & \textbf{Acc} & \textbf{Pre} & \textbf{Recall} & \textbf{F1-Score} \\
\midrule
\text{1cnn1lstm}     & 0.8780  & 0.7875 & 0.7801 & 0.7817 \\
\text{2cnn1lstm}     & 0.8788 & 0.7942 & 0.7881 & 0.7821 \\
\text{2cnn2lstm}     & 0.8777 & 0.7817 & 0.7771 & 0.7772 \\
\text{3cnn1lstm}     & 0.8784 & 0.7890  & 0.7836 & 0.7840  \\
\text{2cnn1Bilstm}   & 0.8781 & 0.7855 & 0.7807 & 0.7805 \\
\text{TCN}           & 0.8918 & 0.8237 & 0.8180 & 0.8179 \\
\bottomrule
\end{tabular}
}
\label{tab:model_comparison}
\end{table}

\section{ Performance Evaluation and Analysis of Results}
\label{sec:results}
This section presents the numerical and graphical results of the proposed FedSecureFormer architecture. It includes an evaluation of attack detection performance, comparisons with Transformer, TCN, and hybrid CNN LSTM models, and an analysis of design choices such as encoder layers, attention pooling, and attention heads. We also validated the FL with DP with a setup of 20 clients and demonstrated real-time inference on a Jetson Nano, confirming the model’s suitability for resource-constrained environments.

\subsection{Models Implemented to Perform IDS}
We experimented with multiple models for attack detection:
\begin{itemize}
\item Hybrid CNN-LSTM: Models trained with varying numbers of CNN, LSTM, and BiLSTM layers with different loss functions.
\item TCN: A basic Temporal Convolutional Network (TCN) model updated using cross-entropy loss.
\item Transformers: Multiple transformer architectures explored by varying the number of layers, multi-head self-attention heads, and employing different loss functions.
\end{itemize}

\subsection{Evaluation and Analysis of Results Obtained by Different Hybrid CNN-LSTM and TCN Models}

Table \ref{tab:model_comparison} presents the performance metrics of the hybrid CNN-LSTM and TCN models. To ensure fairness, all the compared models discussed were evaluated under identical experimental settings. Among the hybrid configurations tested, the 2CNN-1LSTM (2cnn1lstm) architecture—comprising two convolutional layers followed by one LSTM layer achieved the best results when optimized using label smoothing loss. While other loss functions, such as focal and cross-entropy, were explored, label smoothing provided superior performance. Increasing the number of CNN or LSTM layers did not yield noticeable improvements in accuracy, but it significantly increased training time. The TCN model, however, delivered better overall performance due to its strength in modelling temporal dependencies. For clarity, only 2cnn1lstm and TCN models are compared in Table \ref{tab:cnnvstcn}. The TCN achieved an overall recall of 81.79\%, outperforming 2cnn1lstm, which achieved a recall of 78.81\%. 

\begin{table}
\centering
\caption{Recall Comparison Between Hybrid CNN-LSTM and TCN Across Attack Type}
\label{tab:cnnvstcn}
\renewcommand{\arraystretch}{1.2}
\resizebox{0.6\columnwidth}{!}{%
\begin{tabular}{ccc}
\toprule
\textbf{Attack Class} & \textbf{Hybrid CNN-LSTM} & \textbf{TCN} \\
\midrule
A(0)  & 0.5099 & 0.5935 \\
A(1)  & 0.8250 & 0.8697 \\
A(2)  & 0.7300 & 0.7647 \\
A(3)  & 0.9983 & 0.9967 \\
A(4)  & 0.5917 & 0.9468 \\
A(5)  & 0.9717 & 0.9559 \\
A(6)  & 0.9617 & 0.9633 \\
A(7)  & 0.9983 & 0.9942 \\
A(8)  & 0.9983 & 0.9942 \\
A(9)  & 0.8583 & 0.8529 \\
A(10) & 0.7917 & 0.8304 \\
A(11) & 0.4667 & 0.4128 \\
A(12) & 0.6333 & 0.7052 \\
A(13) & 0.9617 & 0.9739 \\
A(14) & 0.7400 & 0.7871 \\
A(15) & 0.5517 & 0.6623 \\
A(16) & 0.9733 & 0.9799 \\
A(17) & 0.4083 & 0.5186 \\
A(18) & 0.9183 & 0.8076 \\
A(19) & 0.8650 & 0.7498 \\
\midrule
\textbf{Overall} & \textbf{0.7881} & \textbf{0.8179} \\
\bottomrule
\end{tabular}}
\end{table}  

\begin{table}
\centering
\caption{Transformer Architecture Evaluation Across Pooling, Encoder and Attention Variants}
\footnotesize
\resizebox{1\columnwidth}{!}{
\renewcommand{\arraystretch}{1.2}
\begin{tabular}{c c c c c c c}
\toprule
\textbf{Model} & \textbf{Pooling used} & \textbf{No of Enc/attn} & \textbf{Acc} & \textbf{Pre} & \textbf{Recall} & \textbf{F1-Score} \\
\midrule
\multirow{8}{*}{Transformer} 
& Mean Average       & 6 layer, 2 attn       & 0.9306 & 0.8786 & 0.8190 & 0.8392 \\
& Single-head attn   & 6 layer, 2 attn       & 0.9304 & 0.8756 & 0.8180 & 0.8382 \\
&                    & 4 layer, 2 attn       & 0.9245 & 0.8671 & 0.8035 & 0.8273 \\
&                    & 4 layer, 1 attn       & 0.9238 & 0.8626 & 0.8039 & 0.8260 \\
& Multi-head attn    & 4 layer, 2 attn 
&0.9245	&0.867 &	0.8035 &	0.827
 \\
&                    & 6 layer, 1 attn       & 0.9306 & 0.8761 & 0.8187 & 0.8380 \\
&                    & \textbf{6 layer, 2 attn} & \textbf{0.9369} & \textbf{0.8798} & \textbf{0.8205} & \textbf{0.8404} \\
\bottomrule
\end{tabular}}
\label{tab:transformer_variants}
\end{table}

\subsection{Evaluation and Analysis of Results obtained by Varying Transformer Architectures}

\begin{figure}
    \centering
    \includegraphics[width=0.4\textwidth]{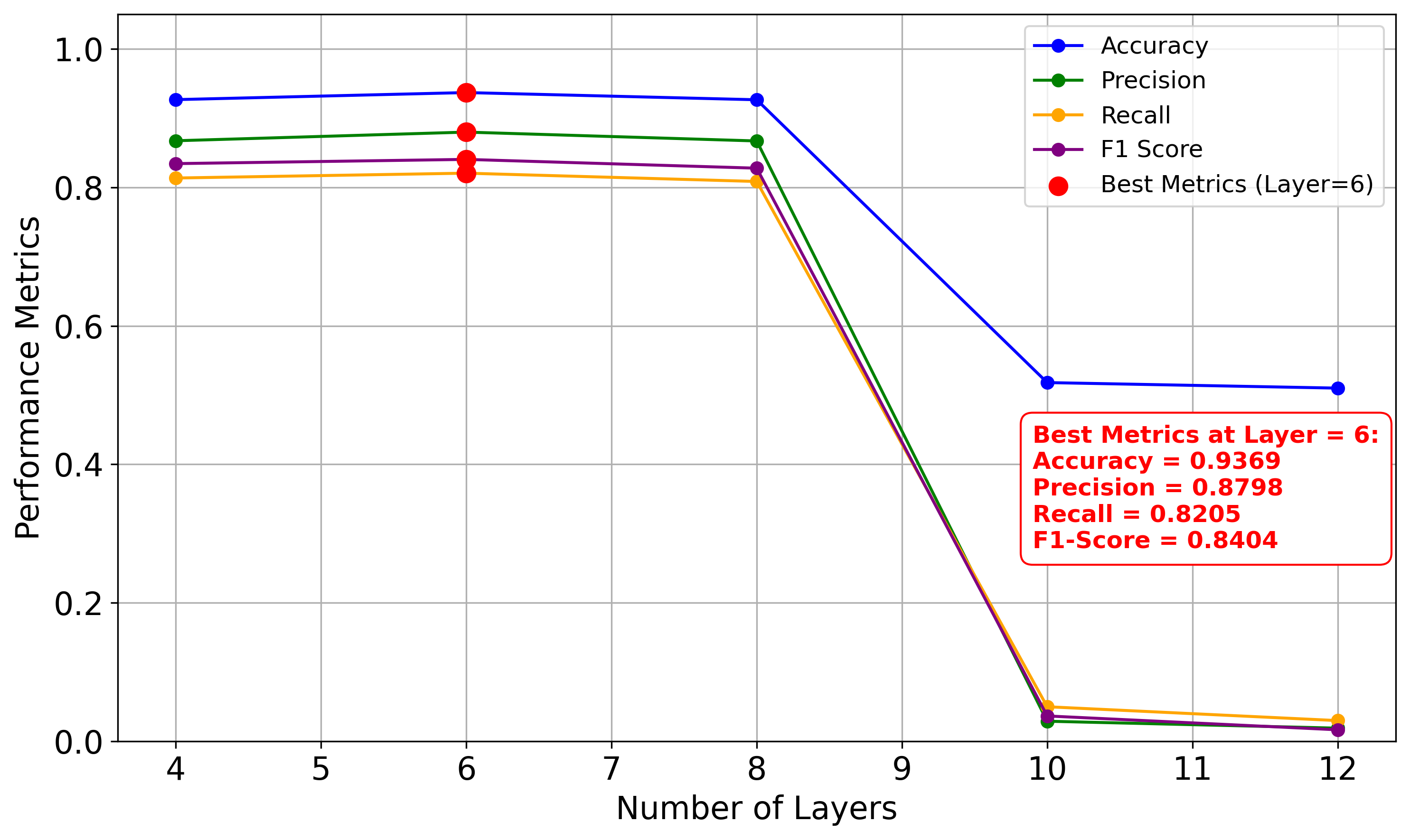}
    \caption{FedSecureFormer achieves optimal performance across Accuracy, Precision, Recall\textcolor{blue}{,} and F1-score plotted against the number of encoder layers. Red labels indicate the best values at six encoder layers.}
    \label{fig:layer_comp}
\end{figure}

\begin{figure}
    \centering
    \includegraphics[width=0.4\textwidth]{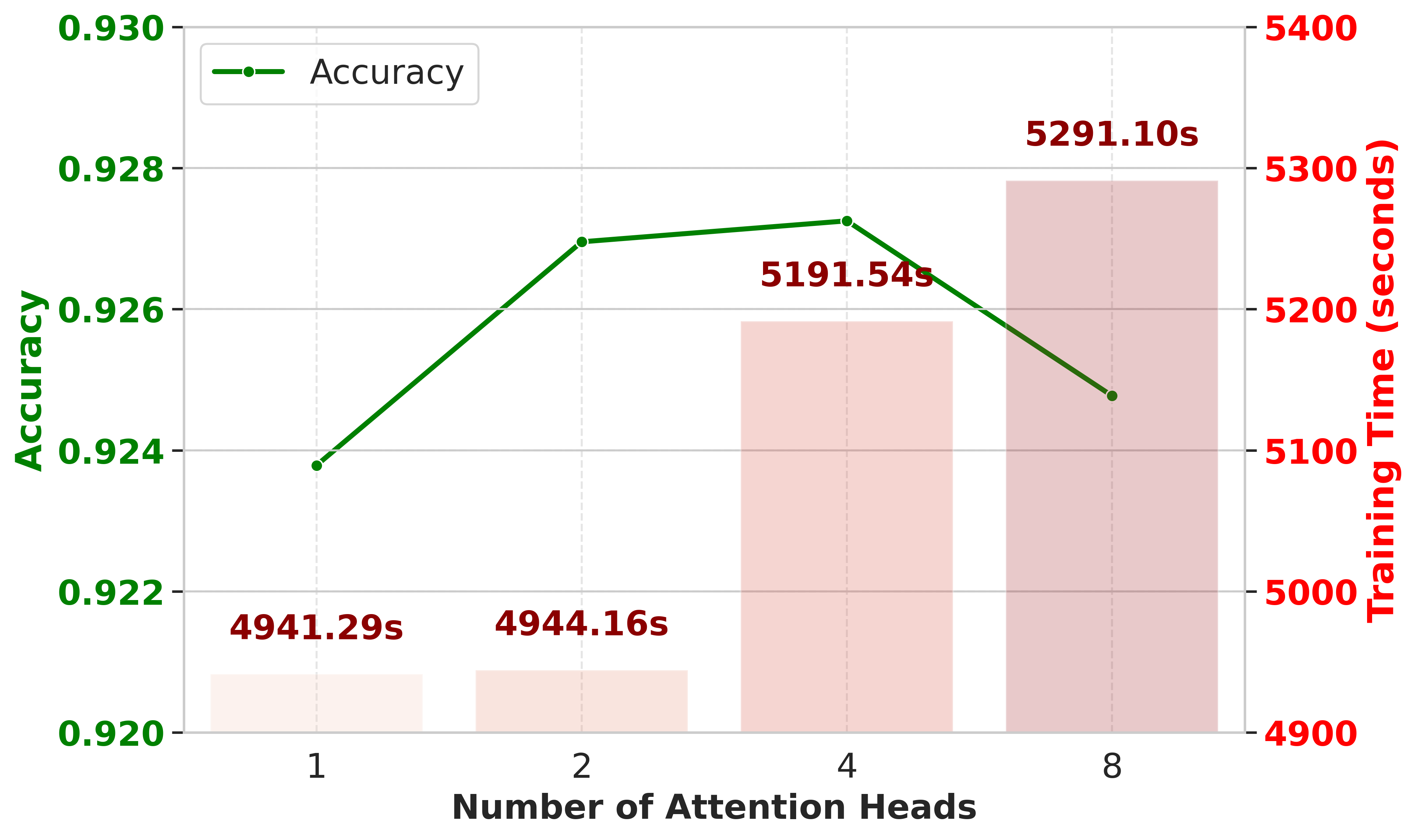}
    \caption{FedSecureFormer achieves similar accuracy with both 2 and 4 attention heads, though 4 heads significantly increase training time.}
    \label{fig:atnn_comp}
\end{figure}

As shown in Table \ref{tab:transformer_variants}, the six-layer transformer encoder with two attention heads and multi-head attention pooling yielded the best results. Multi-head pooling outperformed mean average and single-head pooling in handling time series data. As illustrated in Fig. \ref{fig:layer_comp}, six layers offered the best trade-off between accuracy and complexity, while deeper models led to overfitting. Although four attention heads performed slightly better, as illustrated in Fig. \ref{fig:atnn_comp}, we used two attention heads to reduce cost and retained four in the pooling layer for optimal performance. As reported in Table \ref{tab:trans_recall}, the model achieved an accuracy of 93.69\%, precision of 87.98\%, recall of 82.05\%, and F1 score of 84.04\%. It performed exceptionally well in ten attack types, with many achieving recall above 0.95 and perfect scores in eight attack classes. We used a projection dimension of 64, which reduced model complexity and training time with only a 0.305\% drop in performance compared to the larger variant with a 128 projection dimension.

\captionsetup[figure]{skip=0.5pt}

\subsection{Comparison of Performance with SOTA}

The recall comparison across attack classes illustrated in Table \ref{tab:recall-comparison} shows that our proposed model consistently outperforms traditional baselines and recent SOTA methods with respect to detection capability and reliability. It achieves the highest recall in 10 out of 19 attack classes, including perfect recall (100\%) for critical classes such as A(3), A(7), A(8), and A(18). In contrast, models like Hybrid CNN-LSTM and prior works \cite{devika2024vadgan}, \cite{chougule2022multibranch}, \cite{slama2022comparative} exhibit lower and inconsistent performance. For example, \cite{chougule2022multibranch} records only 5.5, 8, and 9\% recall for classes A(2), A(4), and A(9), respectively. While \cite{slama2022comparative} performs well on A(3), A(7), A(11), and A(12), it lacks consistency across other classes. Our model achieves 99.89\% on A(1), 92.38\% on A(2), 97.25\% on A(5), and 99.94\% on A(19), outperforming all baselines.
\begin{table}[!htbp]
\centering
\caption{Evaluation Metrics per Attack Type Utilizing Our
Proposed Transformer}
\label{tab:trans_recall}
\renewcommand{\arraystretch}{1.2}
\resizebox{0.7\columnwidth}{!}{
\begin{tabular}{ccccc}
\toprule
\textbf{Attack Class} & \textbf{Accuracy} & \textbf{Precision} & \textbf{Recall} & \textbf{F1 Score} \\
\midrule
A(0)  & 0.9691 & 0.9694 & 0.9997 & 0.9843 \\
A(1)  & 0.8295 & 0.9122 & 0.9514 & 0.9068 \\
A(2)  & 0.9581 & 0.9038 & 0.6192 & 0.7349 \\
A(3)  & 0.9935 & 0.9935 & 1.0000 & 0.9967 \\
A(4)  & 0.8798 & 0.9765 & 0.8988 & 0.9361 \\
A(5)  & 0.9715 & 0.9835 & 0.9876 & 0.9855 \\
A(6)  & 0.9656 & 0.9896 & 0.9896 & 0.9825 \\
A(7)  & 0.9973 & 0.9973 & 1.0000 & 0.9986 \\
A(8)  & 0.9960 & 0.9960 & 1.0000 & 0.9980 \\
A(9)  & 0.9509 & 0.9534 & 0.9214 & 0.6742 \\
A(10) & 0.9742 & 0.9048 & 0.8042 & 0.8515 \\
A(11) & 0.9318 & 0.4432 & 0.5297 & 0.4826 \\
A(12) & 0.9588 & 0.9977 & 0.5888 & 0.7406 \\
A(13) & 0.9385 & 0.9427 & 0.9952 & 0.9683 \\
A(14) & 0.9737 & 0.9837 & 0.7464 & 0.8488 \\
A(15) & 0.6348 & 0.8173 & 0.7397 & 0.7766 \\
A(16) & 0.9719 & 0.9943 & 0.9773 & 0.9857 \\
A(17) & 0.9218 & 0.4029 & 0.3209 & 0.3572 \\
A(18) & 0.9717 & 0.7238 & 1.0000 & 0.8352 \\
A(19) & 0.9514 & 0.6477 & 0.7133 & 0.6789 \\
\midrule
\textbf{Overall} & \textbf{0.9369} & \textbf{0.8798} & \textbf{0.8205} & \textbf{0.8404} \\
\bottomrule
\end{tabular}}
\end{table}

\begin{table}[t]
\centering
\caption{Comparison of Recall Rates for Various Attack Types with SOTA}
\label{tab:recall-comparison}
\resizebox{0.95\columnwidth}{!}{%
\begin{tabular}{@{}c|ccccc@{}}
\toprule
\textbf{Attack Class} & \textbf{Our Model} & \textbf{Hybrid CNN-LSTM} & \textbf{\cite{devika2024vadgan}} & \textbf{\cite{chougule2022multibranch}} & \textbf{\cite{slama2022comparative}} \\
\midrule
A(1)  & \textbf{0.9514} & 0.8250 & 0.9100 & 0.4250 & 0.8234 \\
A(2)  & 0.6192          & 0.7300 & 0.4200 & 0.0550 & 0.7467 \\
A(3)  & \textbf{1.0000} & 0.9983 & 0.7500 & 1.0000 & 0.9930 \\
A(4)  & 0.8988          & 0.5917 & 0.4200 & 0.0800 & 0.9936 \\
A(5)  & \textbf{0.9876} & 0.9717 & 0.9425 & 0.9450 & 0.8181 \\
A(6)  & \textbf{0.9896} & 0.9617 & 0.9860 & 0.4200 & 0.0381 \\
A(7)  & \textbf{1.0000} & 0.9983 & 0.9000 & 1.0000 & 0.9948 \\
A(8)  & \textbf{1.0000} & 0.9983 & 0.9135 & 1.0000 & 0.9855 \\
A(9)  & \textbf{0.9214} & 0.8583 & 0.9000 & 0.0900 & 0.0653 \\
A(10) & 0.8042          & 0.7917 & 1.0000 & 0.9850 & 0.0681 \\
A(11) & 0.5297          & 0.4667 & 1.0000 & 0.9850 & 0.9992 \\
A(12) & 0.5888          & 0.6333 & 0.6000 & 0.0850 & 0.9990 \\
A(13) & \textbf{0.9952} & 0.9617 & 1.0000 & 0.9800 & 0.7825 \\
A(14) & 0.7464          & 0.7400 & 1.0000 & 1.0000 & 0.7888 \\
A(15) & 0.7397          & 0.5517 &  0.4800      & 0.9950 & 0.6004 \\
A(16) & \textbf{0.9773} & 0.9733 & 0.8850 & 0.9850 & 0.0278 \\
A(17) & 0.3209          & 0.4083 & 1.0000 & 0.9750 & 0.5326 \\
A(18) & \textbf{1.0000} & 0.9183 & 0.9333 & 1.0000 & 0.7958 \\
A(19) & 0.7133          & 0.8650 & 0.7635 & 0.9900 & 0.7693 \\
\bottomrule
\end{tabular}
}
\end{table}

\begin{table}[t]
\centering
\caption{Comparison of FedAvg Strategy in Different Settings}
\label{tab:fedavg}
\small
\renewcommand{\arraystretch}{1.2}
\resizebox{0.95\columnwidth}{!}{
\begin{tabular}{@{}lccccccc@{}} 
\toprule
\textbf{Strategy} & \textbf{\# Clients} & \textbf{Epochs} & \textbf{Rounds} & \textbf{Acc} & \textbf{Pre} & \textbf{Recall} & \textbf{F1-score} \\
\midrule
Centralized & -- &-- & 100  & 0.9369 & 0.8798 & 0.8205 & 0.8404\\
\midrule
\multirow{3}{*}{FedAvg} & \multirow{3}{*}{20} & 1 & 100 &\textbf{0.8301} & \textbf{0.9094} & \textbf{0.8457} & \textbf{0.8561} \\
 &  & 2 & 50  & 0.8202 & 0.8960 & 0.8202 & 0.8439 \\
 &  & 4 & 25  & 0.7551 & 0.8829 & 0.7551 & 0.7998 \\
\bottomrule
\end{tabular}
}
\end{table}

Even for challenging classes like A(11) and A(12), it maintains a competitive recall of 52.97\%\textcolor{blue}{,} and 58.88\%, respectively, surpassing most benchmarks. Additionally, we also encountered the study \cite{hamhoum2024mistralbsm}, which utilizes the Mistral-7B model for attack detection in CAVs, employing 8 out of 19 attack classes from the VeReMi Extension dataset. However, the paper has reported only class-averaged metrics, with ours having an average improvement of approximately 0.0058 across all performance measures. Since the paper failed to report per-class metrics, a further detailed assessment is not possible.
\subsection{Evaluation of Federated Learning Results}
\begin{figure}[]
    \centering
    \includegraphics[width=0.4\textwidth]{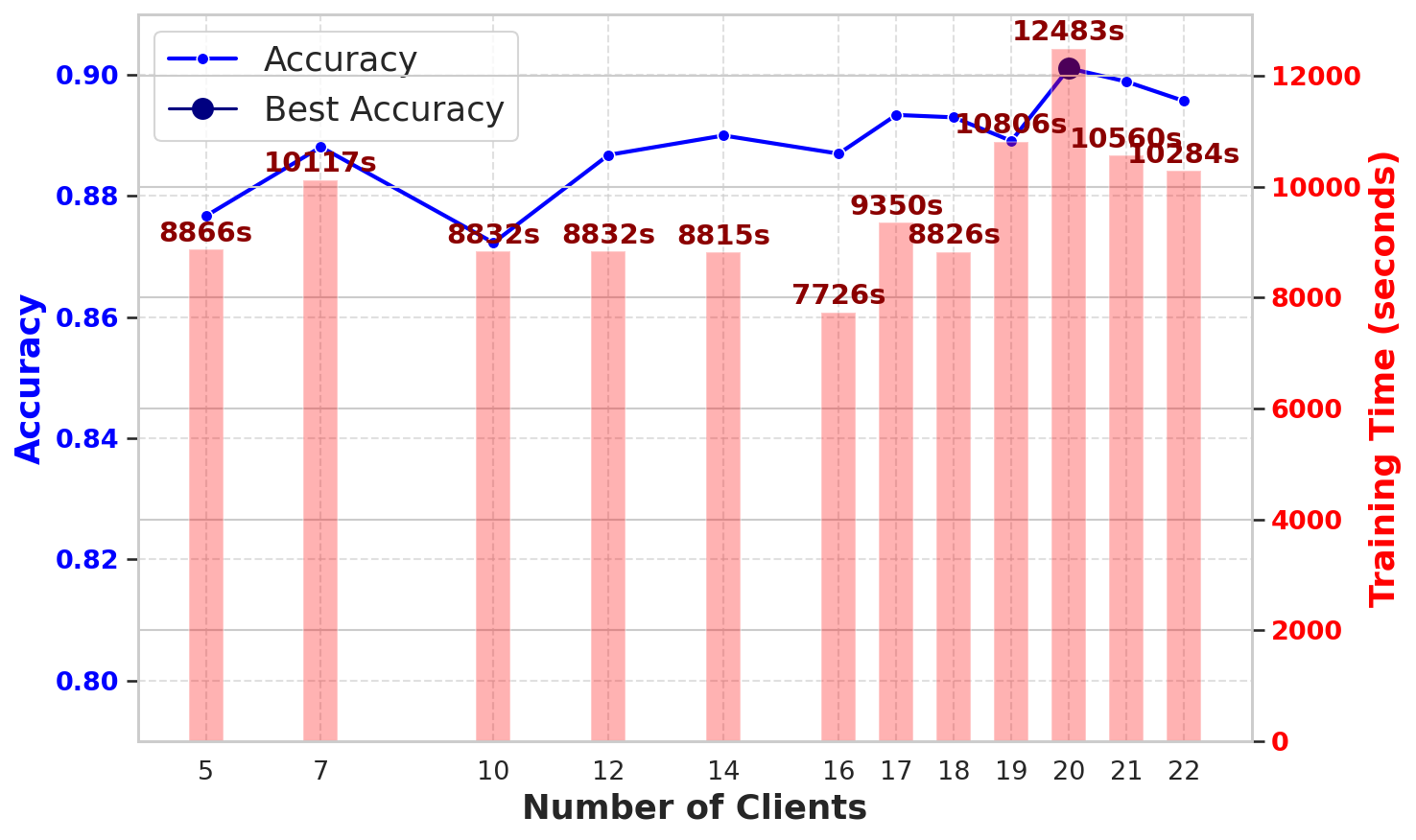}
    \caption{Performance of FedSecureFormer in a federated setup, showcasing the ideal number of clients as 20.}
    \label{fig:fl_20best}
\end{figure}

In the FL setup using FedSecureFormer, the best performance was observed with 20 clients as illustrated in Fig. \ref{fig:fl_20best}. 
\begin{table}[]
\centering
\caption{FedProx Strategy Performance with Varying Clients and Proximal Parameters}
\label{tab:fedprox}
\small
\renewcommand{\arraystretch}{1.2}
\resizebox{1.0\columnwidth}{!}{
\begin{tabular}{@{}llcccccccc@{}}
\toprule
\textbf{Strategy} & \textbf{\# Clients} & \textbf{Epochs} & \textbf{Rounds} & \textbf{Acc} & \textbf{Pre} & \textbf{Recall} & \textbf{F1}  & \textbf{$\mu$} \\
\midrule
\multirow{4}{*}{FedProx} 
& \multirow{2}{*}{5} & \multirow{4}{*}{10} & \multirow{4}{*}{50}
& 0.7841 & 0.6761 & 0.6761 & 0.7499  & 0.01 \\
& & & & 0.6652 & 0.3798 & 0.3798 & 0.4150 & 0.1 \\
& \multirow{2}{*}{7} & & & 0.7541 & 0.4027 & 0.4027 & 0.4348  & 0.01 \\
& & & & 0.6589 & 0.3201 & 0.3201 & 0.3086  & 0.1 \\
\bottomrule
\end{tabular}
}
\end{table}
Each client received approximately 11350 training and 2830 test samples, with slight variations to simulate real-world traffic scenarios. Clients participated in a first-come, first-served manner. We used equivalent communication rounds (100) and local epochs (1) to align with centralized training. The FedAvg strategy yielded strong results with only a 1.03\% performance drop compared to centralized training\textcolor{blue}{,} as illustrated in Table \ref{tab:fedavg}. We also tested FedProx but observed inferior performance. As shown in Table \ref{tab:fedprox}, lower values of the proximal term \textbf{$\mu$} yielded better performance.
 Additionally, results for 5 and 7 clients were evaluated using 10 local epochs and 50 communication rounds. Among these, the best performance was obtained with  5 clients, when executed with a \textbf{$\mu$} value of 0.01. We have only shown the results for 5 and 7 clients, which performed best among a subset of configurations chosen from a range of 3 to 10.

\subsection{Federated Learning Executed under Differential Privacy}
Table \ref{tab:fl-dp} compares the performance of FL models with DP under different noise and clipping multipliers, using a batch size of 64. Performance dropped significantly when the noise multiplier exceeded 0.5, indicating sensitivity to noise. However, careful tuning enabled a balance between privacy and accuracy.
\captionsetup[figure]{skip=0.5pt}
\begin{figure*}
    \centering
    \includegraphics[width=1.0\textwidth]{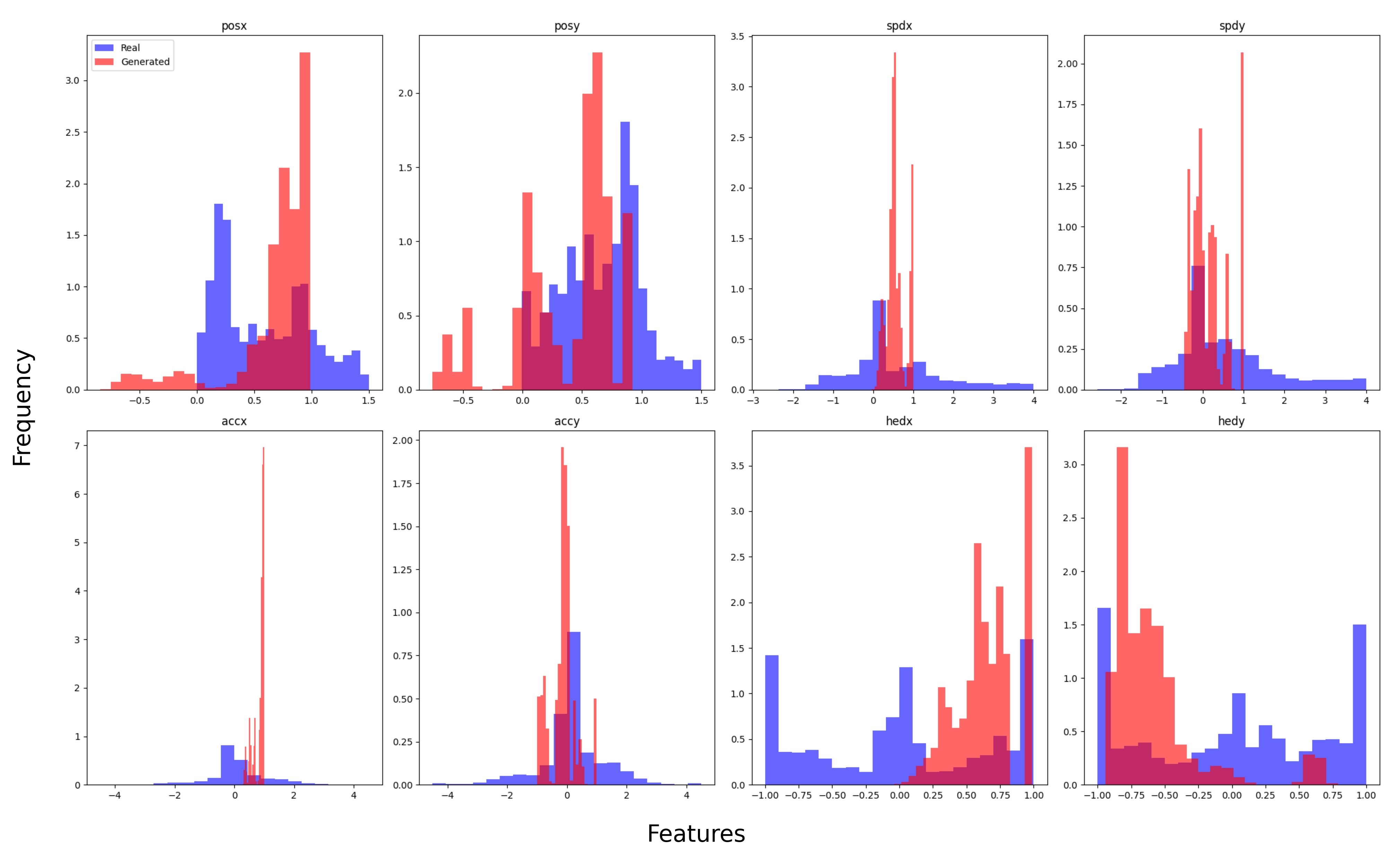}
    \caption{Comparison of real and generated sequences plotted using histograms. Each subplot corresponds to one feature, with the x-axis representing feature values and the y-axis showing frequency counts.}
    \label{fig:real_gen}
\end{figure*}
\begin{table}
\centering
\caption{Evaluation of Federated Learning under Various Differential Privacy Settings}
\label{tab:fl-dp}
\small
\renewcommand{\arraystretch}{1.2}
\resizebox{0.9\columnwidth}{!}{
\begin{tabular}{lccccccc}
\toprule
\textbf{Strategy} & \textbf{Noise} & \textbf{Clip} & \textbf{Acc} & \textbf{Pre} & \textbf{Recall} & \textbf{F1} \\
& \textbf{Multiplier} & \textbf{Norm} & & & & \\
\midrule
\multirow{7}{*}{FL with DP} 
& 0.1   & 1 & 0.441 & 0.791 & 0.441 & 0.501 \\
& 0.05  & 1 & 0.478 & 0.804 & 0.478 & 0.543 \\
& 0.5   & 1 & 0.278 & 0.635 & 0.278 & 0.320 \\
& 0.01  & 3 & 0.565 & 0.870 & 0.565 & 0.644 \\
& 0.001 & 1 & 0.731 & 0.907 & 0.731 & 0.789 \\
& 0.001 & 3 & 0.679 & 0.907 & 0.679 & 0.749 \\
& \textbf{0.001} & \textbf{5} & \textbf{0.827} & \textbf{0.898} & \textbf{0.749} & \textbf{0.797} \\
\bottomrule
\end{tabular}}
\end{table}
The optimal DP-FL setting (noise = 0.001, clipping = 5) achieved an accuracy of 0.827, precision of 0.898, recall of 0.749, and an F1 score of 0.797, closely matching the non-private FL results with only a 4.04\% performance loss. The privacy budget parameters $(\epsilon, \delta)$ achieved with our model are (6.329, 1.00e-05). This shows that well-tuned DP-FL models can preserve privacy with minimal accuracy trade-off.

\subsection{Detection of Unseen Attacks Using Hist-AttnGAN}

The real and generated sequences are presented in Fig.
\ref{fig:real_gen}, where the x-axis corresponds to the feature values across
eight dynamic vehicle features and the y-axis represents
the frequency of observations of each feature. The generated sequence exhibits narrow distributions, as our goal was not to introduce diversity but to create a new class of previously unseen data.
 The generated sequences were fed to the FedSecureFormer Transformer. We evaluated similarity thresholds ranging from 0.3 to 0.9 based on model confidence. We observed underfitting at thresholds below 0.5 and overfitting at thresholds above 0.5, making 0.5 the optimal
similarity threshold. Using this optimal threshold of 0.5, we achieved a detection accuracy of 88\%.

\subsection{Comparison of Model Parameters with Encoder-Only Transformers}
\begin{figure}
    \centering
    \includegraphics[width=0.5\textwidth]{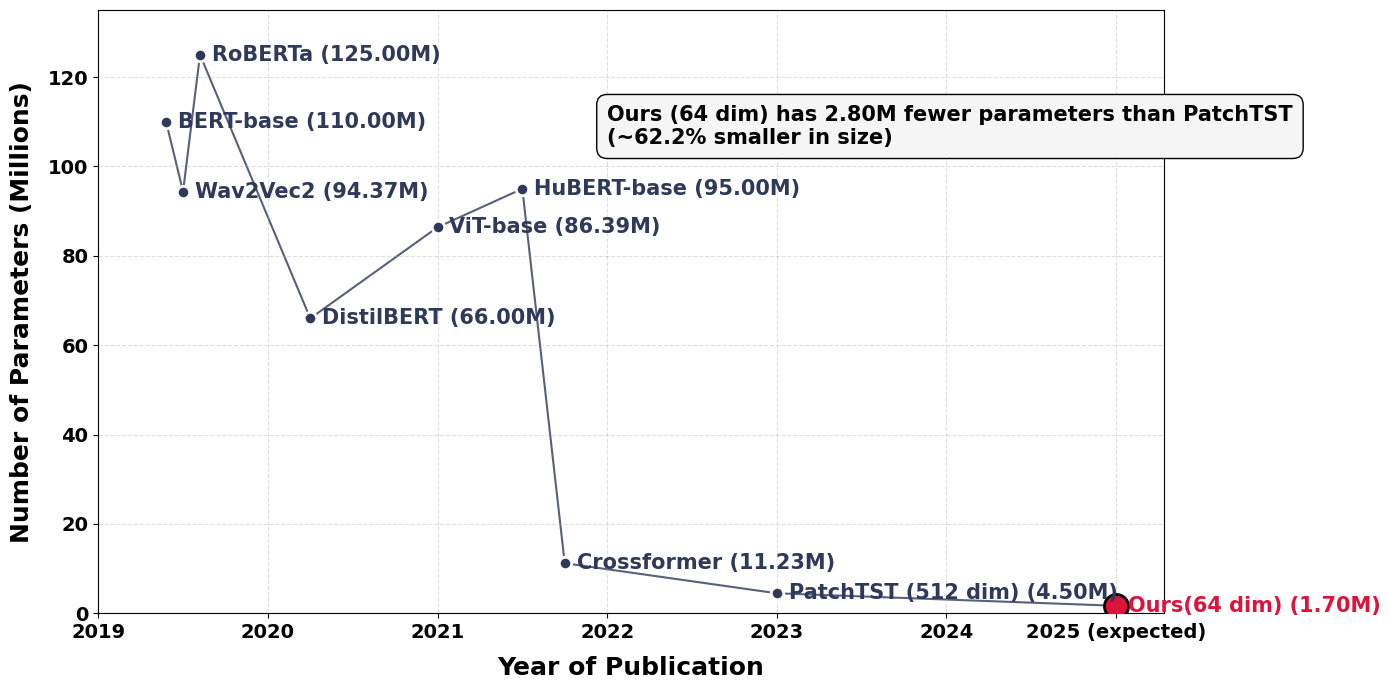}
    \caption{Comparison of FedSecureFormer model parameters with other Encoder-only Transformers.}
    \label{fig:model_param_comp}
\end{figure}

As illustrated in Fig. \ref{fig:model_param_comp}, the proposed FedSecureFormer has 1.70 million parameters, significantly fewer than those of all other compared encoder-only transformer models. The minimal architecture of our model features six encoder layers, two attention heads per encoder, multi-query multi-head attention pooling with four attention heads, a projection layer of 64 dimensions, trained on short sequences and dedicated for time series classification tasks, making it compact with fewer parameters. 

\subsection{Comparison of Inference Time with SOTA}
Table \ref{tab:inference_time} illustrates the deployment of FedSecureFormer on a Jetson Nano for inference to demonstrate real-time applicability. The model achieved an average inference time of 3.7775 milliseconds per vehicle, highlighting its suitability for resource-constrained edge environments. Notably, our model outperforms existing SOTA approaches in terms of inference speedup. Specifically, FedSecureFormer achieves a speedup of 135.53× compared to \cite{chougule2022multibranch} and 66×  compared to \cite{alladi2023deep} when evaluated on the Jetson Nano hardware platform.

\begin{table}
    \centering
    \caption{Inference Time Comparison on Jetson Nano with SOTA}
    \renewcommand{\arraystretch}{1.2} 
    \setlength{\tabcolsep}{6pt} 
    \resizebox{0.5\textwidth}{!}{
    \begin{tabular}{c|c|c|c|c}
        \toprule
        \textbf{Reference} & \textbf{Environment} & \textbf{Data Setup} & \textbf{Prediction} & \textbf{Inference Time (ms)} \\
        \midrule
        \cite{chougule2022multibranch} & \multirow{3}{*}{Jetson Nano} & 0.07 & 511.82 & 511.89 \\
        \cite{alladi2023deep} &  & 0.33  &287.74  & 288.07 \\
        \cite{alladi2023deep} &  &  0.32& 251.24 & 251.56 \\
        Ours &  & \textbf{0.0075} & \textbf{3.77}  & \textbf{3.7775} \\
        
        \bottomrule
    \end{tabular}
    }
    \label{tab:inference_time}
\end{table}

\section{Conclusion}
\label{sec:conc}
In this work, we propose FedSecureFormer, a novel lightweight transformer model for cyberattack detection in Connected and Autonomous Vehicles (CAVs). The model features a six-layer encoder-only architecture with two attention heads, multi-query multi-head attention pooling, and a 64-dimensional projection. Comparable performance was observed between 64 and 128 dimensions, with the 64-dimensional version reducing the model size to 1.7 million parameters (approximately 1.9M fewer than the larger variant). Despite its compact design, the model achieved a classification accuracy of 93.69\% across 19 attack categories. FedSecureFormer demonstrated strong resilience in real-world scenarios, with only a 1.03\% accuracy drop in the federated learning setup and a 4.04\% performance loss under differential privacy. It also generalized well to unseen adversarial attacks generated using a histogram-based LSTM attention GAN, achieving 88\% detection accuracy. Most importantly, it delivered real-time inference with an average of 3.7775 milliseconds per vehicle on a Jetson Nano, making it nearly 100 times faster than existing state-of-the-art models and ideal for latency-sensitive Intelligent Transport Systems. We aim to explore the use of pre-trained transformer models to benchmark performance against our custom architecture. We also plan to investigate more efficient federated learning strategies that are mindful of bandwidth limitations, under heterogeneous data distributions, minimize the performance gap between centralized and FL settings, and are better suited for deployment on edge devices.

\bibliographystyle{IEEEtranN}
{\footnotesize
\bibliography{ref}}

\end{document}